\documentclass[runningheads]{llncs}
\usepackage[T1]{fontenc}
\usepackage{graphicx,verbatim}
\usepackage{comment}
\usepackage{tabularx}
\usepackage{booktabs}
\usepackage{amsmath}
\usepackage[colorlinks=true, linkcolor=blue, citecolor=blue, urlcolor=blue]{hyperref}
\usepackage{float}
\usepackage{arydshln}  

\begin{document}
%



\title{Comparing Conditional Diffusion Models for Synthesizing Contrast-Enhanced Breast MRI from Pre-Contrast Images}

\titlerunning{Comparing Conditional Diffusion Models for Synthesizing Breast MRI}
\author{Sebastian Ibarra\inst{1}
\and
Javier del Riego\inst{2}
\and
Alessandro Catanese\inst{3}
\and
Julian Cuba\inst{4}
\and
Julian Cardona\inst{4}
\and
Nataly Leon\inst{4}
\and
Jonathan Infante\inst{4}
\and
Karim Lekadir\inst{1,5}
\and
Oliver Diaz\inst{1,6}
\and
Richard Osuala\inst{1}
}
\authorrunning{S. Ibarra et al.}

\institute{Departament de Matemàtiques i Informàtica, Universitat de Barcelona, Spain
\email{sebastian.ibarra@ub.edu, richard.osuala@ub.edu}
\and
Parc Taulí Hospital Universitari, Spain
\and
Hospital Germans Trias i Pujol, Spain
\and
Siemens Healthineers, America, LAM
\and
Institució Catalana de Recerca i Estudis Avançats (ICREA), Barcelona, Spain
\and
Computer Vision Center, Bellaterra, Spain
}

\maketitle              
\begin{abstract}
Dynamic contrast-enhanced (DCE) MRI is essential for breast cancer diagnosis and treatment. However, its reliance on contrast agents introduces safety concerns, contraindications, increased cost, and workflow complexity.
To this end, we present pre-contrast conditioned denoising diffusion probabilistic models to synthesize 
DCE-MRI, 
introducing, evaluating, and comparing a total of 22 generative model variants in both single-breast and full breast settings. 
Towards enhancing lesion fidelity, we introduce both tumor-aware loss functions and explicit tumor segmentation mask conditioning.
Using a public multicenter 
dataset and comparing to respective pre-contrast baselines, 
we observe that subtraction-image-based models consistently outperform post-contrast-based models across five complementary evaluation metrics.
Apart from assessing the entire image, we also separately evaluate the region of interest, where both tumor-aware losses and segmentation mask inputs improve evaluation metrics. The latter notably enhance qualitative results capturing contrast uptake, albeit assuming access to tumor localization inputs that are not guaranteed to be available in screening settings.
A reader study involving 2 radiologists and 4 MRI technologists confirms the high realism of the synthetic images, indicating an emerging clinical potential of generative contrast-enhancement.
We share our codebase at 
\url{https://github.com/sebastibar/conditional-diffusion-breast-MRI}.



\end{abstract}

\section{Introduction}

Breast cancer is the most frequently diagnosed cancer in women, with 2.3 million new cases and 670,000 deaths in 2022~\cite{bray2024global}. Early detection improves outcomes, and breast magnetic resonance imaging (MRI) offers high sensitivity for screening high-risk individuals~\cite{lehman2005screening}. In this regard, dynamic contrast-enhanced MRI (DCE-MRI) is the clinical gold standard, leveraging gadolinium-based contrast agents (GBCAs) to localize and visualize tumor vascularity~\cite{mann2019breast}.
Despite its high diagnostic value, DCE-MRI is time-consuming and expensive as well as
contraindicated in vulnerable populations~\cite{chhor2017abbreviated,osuala2025simulating}. GBCAs further pose risks such as nephrogenic systemic fibrosis~\cite{sardanelli2020gadolinium}, gadolinium retention~\cite{xie2022magnetic}, allergic-like reactions as well as other Symptoms Associated with Gadolinium Exposure (SAGE)\cite{mcdonald2022symptoms}.
These limitations have driven interest in contrast-free imaging alternatives. Deep generative models—particularly conditional architectures—can synthesize post-contrast images from non-contrast inputs, with promising early results~\cite{denck2021enhanced,muller2024diffusion,osuala2025simulating,osuala2024towards,lang2025temporalneuralcellularautomata,schreiter2024virtual}. 
With other generative modeling techniques
often suffering from instability, low generalization, or poor perceptual quality~\cite{dimitriadis2022enhancing}, image-level denoising diffusion probabilistic models~\cite{ho2020denoising} (DDPMs) offer a compelling alternative via iterative denoising, enabling stable, high-fidelity synthesis~\cite{song2021score}. Recent studies demonstrate their effectiveness in 
medical image denoising and anomaly detection, as well contrast-agent dose reduction~\cite{muller2024diffusion}, and segmentation-guided anatomy generation in breast MRI~\cite{konz2024anatomically}. Although latent diffusion models \cite{rombach2022high} have been recently introduced for pre- to post-contrast breast MRI synthesis \cite{osuala2024towards}, their image-level counterparts \cite{muller2024diffusion} are still an underexplored method in this domain, e.g., missing comprehensive benchmarking across multiple model variants and preprocessing methods, supervision strategies, and clinical settings.
To this end, the present study 
analyses the potential of DDPMs for realistic, contrast-agent-free DCE-MRI synthesis including the following contributions: 
\begin{itemize}
    \item Implementation and benchmarking framework for pre-contrast-conditioned DDPMs for DCE-MRI synthesis.
    \item Proposal of tumor-aware losses 
    and segmentation-guided model 
    conditioning.
    \item Comparison between post-contrast and subtraction image targets, as well as full and single-breast synthesis strategies. 
    \item Expert-reader study alongside quantitative and qualitative lesion and image-level assessment demonstrating synthesis realism, particularly in subtraction-based and tumor-aware models.
\end{itemize}

\section{Material and Methods}
\label{sec:materialmethods}

\subsection{Dataset and Preprocessing}

\begin{figure}[ht]
\centering
\includegraphics[width=0.92\textwidth]{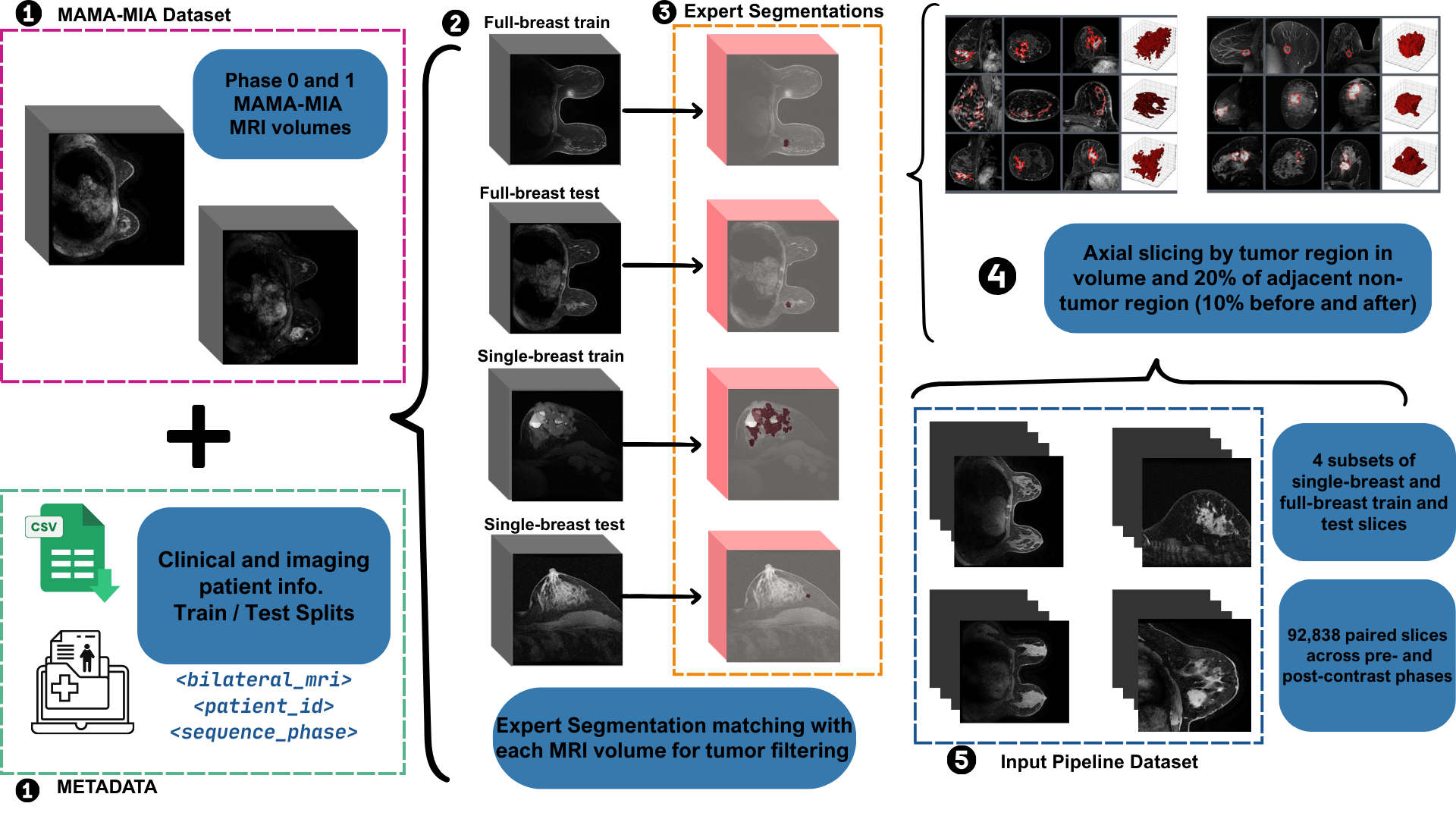}
\caption{ Preprocessing workflow from MRI volumes to the final tumor-containing single-breast and full-breast slice datasets used to train and evaluate our diffusion models.}
\label{fig:preprocessing}
\end{figure}

We use the MAMA-MIA dataset~\cite{garrucho2025}, a publicly available large-scale heterogeneous breast DCE-MRI benchmark, spanning 1,506 multi-source pre-treatment DCE-MRI cases, designed to support algorithm development for tumor segmentation, treatment response prediction, and contrast synthesis. 
%
%
For each patient, we selected Phase~0 (pre-contrast) as model input and Phase~1 (early post-contrast) as target, ensuring anatomical alignment and temporal relevance.
Based on MAMA-MIA's expert-corrected tumor segmentation masks, all slices containing tumor voxels were extracted, along with 20\% of adjacent non-tumor slices to increase data diversity.
For slice extraction, volumes were loaded as 3D NIfTI files, sliced axially, and then, each 2D axial slice was min-max normalized to [0,1] based on its own intensity range, then scaled to [0,255] and exported as 8-bit PNG. 
Data were split into four subsets (unilateral/bilateral, train/test), with accompanying metadata including patient ID, slice index, and tumor label. Patients were classified as unilateral or bilateral based on metadata; all images retain the full axial field of view. The final dataset contains 92,838 paired 2D images across contrast phases. A visualization of the preprocessing pipeline is depicted in the Fig. \ref{fig:preprocessing}.

\subsection{Contrast-Enhancing Generative Modeling}

\begin{figure*}[ht]
\centering
\includegraphics[width=1\textwidth]{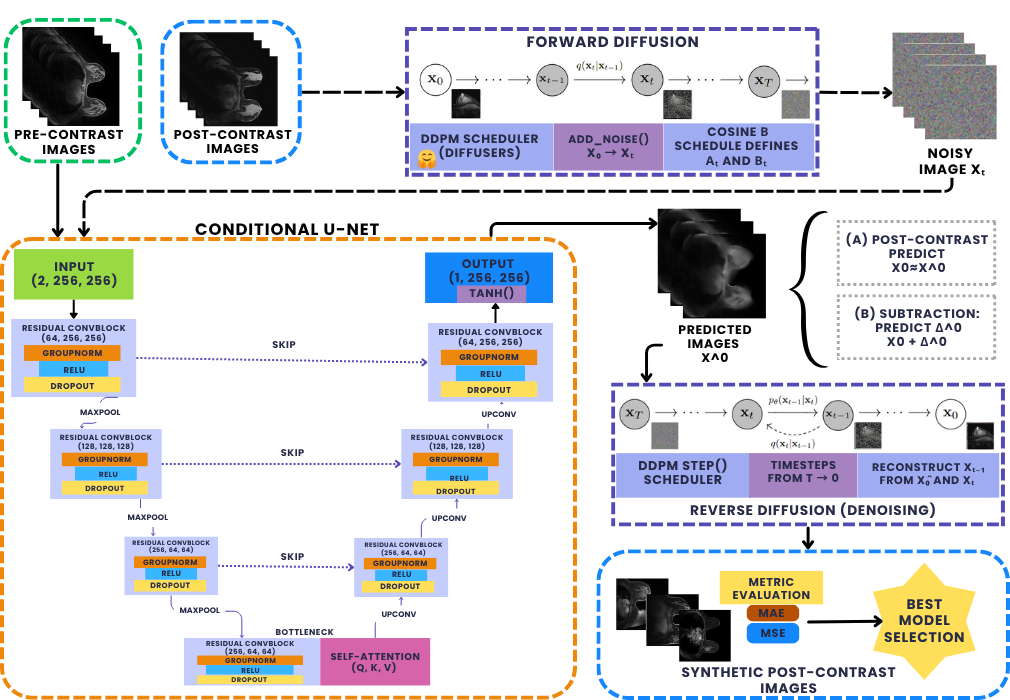}
\caption{Overview depicting both post-contrast and subtraction-based diffusion model training pipelines and model architectures.}
\label{fig:workflow}
\end{figure*}

As illustrated in Figure \ref{fig:workflow}, we develop and compare two conditional DDPMs for synthetic post-contrast MRI generation. Firstly, the post-contrast DDPM directly predicts the full post-contrast image from a respective pre-contrast input image. A second alternative subtraction-based DDPM learns to predict the residual enhancement $(x_{\text{post}} - x_{\text{pre}})$, which is added back to the pre-contrast image to reconstruct the corresponding post-contrast image.

\paragraph{Post-Contrast Conditional DDPM:} \label{sec:loss}
To this end, we implement a conditional U-Net model with residual blocks and a bottleneck self-attention layer, which takes a 2-channel input (pre-contrast + noisy post-contrast at timestep $t$) and predicts the post-contrast image $\hat{x}_0$.
%
Following the image-predictive variant in~\cite{von-platen-etal-2022-diffusers}, the model regresses $\hat{x}_0$ instead of noise $\epsilon$, enforcing pixel-level consistency and improving anatomical fidelity. Gaussian noise is added to $x_0$ using a cosine schedule. With $c$ as the conditioning pre-contrast image, during denoising the model computes the reverse posterior mean according to:
\[
\mu_\theta(x_t, c, t) = \frac{1}{\sqrt{\alpha_t}} \left( x_t - \frac{1 - \alpha_t}{\sqrt{1 - \bar{\alpha}_t}} \hat{x}_0(x_t, c, t) \right)
\]
%
We further define our training objective as a weighted combination of image-level comparison loss functions based on the intuition of a balance of pixel accuracy, structure, and spatial smoothness: 
$\mathcal{L}_{\text{total}} = 
0.3\, \mathcal{L}_{\text{MAE}} + 0.6\, \mathcal{L}_{\text{Perceptual}} + 
0.15\,\mathcal{L}_{\text{Total Variation (TV)\cite{rudin1992nonlinear}}} + 
0.05\,\mathcal{L}_{\text{MSE}}$.
%
To improve convergence efficiency, we adopt the AdamW optimizer \cite{loshchilov2017decoupled}, exponential moving average (EMA) models with $\lambda=0.999$ and mixed-precision training using \cite{accelerate}.

\paragraph{Subtraction-Based Conditional DDPM:}

This variant learns to predict the scaled subtraction image $x_{\text{sub}} = (x_{\text{post}} - x_{\text{pre}})/0.5$, which represents the localized contrast enhancement. During inference, the synthetic post-contrast image is reconstructed as $x_{\text{post}}^{\text{gen}} = 0.5 \cdot \hat{x}_0 + x_{\text{pre}}$, where $\hat{x}_0$ denotes the predicted clean subtraction image.
The architecture, noise schedule, and training setup mirror those of the post-contrast DDPM. However, the loss is computed on the reconstructed post-contrast image to ensure anatomical fidelity and contrast realism.

\subsection{Tumor-Aware Model Training}

We implement two tumor region-of-interest (ROI) aware models. In the first, the expert segmentation mask is concatenated as an additional channel to the input, thereby facilitating the model to focus on learning contrast enhancement patterns within lesion-containing regions of the pre-contrast images.
%
%
Instead of modifying the model input, the second ROI-aware variant receives the full pre-contrast image but introduces lesion-specific supervision through a spatially-weighted loss guided by the expert segmentation mask $M$. Based on $M$, the ROI loss $\mathcal{L}_{\text{ROI}}$ combines (1) pixel-wise MAE and MSE within the tumor region, (2) a VGG-based perceptual loss over the tumor region, and (3) a total variation (TV) \cite{rudin1992nonlinear} term to penalize high-frequency noise and to promote spatial smoothness.

Motivated by the intuition of designing a well-rounded set of complementary region-aware losses, and supported by visual observations of underenhancement in tumor regions, we add a contrast-specific MAE term that penalizes only underestimated signals using ReLU-masked residuals: $\mathcal{L}_{\text{MAEcontrast}} = \text{MAE}([\hat{x}_{\text{ROI}} - x_{\text{pre}} \cdot M]_+, [x_{\text{ROI}} - x_{\text{pre}} \cdot M]_+)$. To preserve mean enhancement intensity, we also include an intensity loss term defined as $\mathcal{L}_{\text{intensity}} = |\mu(\hat{x}_{\text{ROI}}) - \mu(x_{\text{ROI}})|$.
The final loss is a weighted sum: $\mathcal{L}_{\text{total}} = 0.3\,\mathcal{L}_{\text{global}} + 0.6\,\mathcal{L}_{\text{ROI}} + 0.05\,\mathcal{L}_{\text{MAEcontrast}} + 0.05\,\mathcal{L}_{\text{intensity}}$, in order to balance tumor-aware fidelity with global accuracy via the $\mathcal{L}_{\text{global}}$ loss computed on the full image as defined in \ref{sec:loss}.


\subsection{Reader Study: Clinical Validation}

Six domain experts evaluated the synthesis realism in a three-part visual assessment, including two radiologists with 11+ and 9+ years practicing experience 
and four expert MRI technologists with 10 to 15+ years of experience. 
In Task 1 (Discrimination), 15 mixed images (10 synthetic, 5 real) are shown individually to assess distinguishability between real and synthetic. In Task 2 (Comparative), readers were shown randomly selected triplets (pre-contrast, real post, synthetic post) in order to identify the real image between the two post-contrast images. In Task 3 (Annotation), experts annotate differences and score realism within triplets already labeled as real and synthetic. A semi-structured follow-up discussion captured further 
insights on diagnostic relevance and clinical integration.

\begin{figure}[ht]
    \centering
    \includegraphics[width=0.9\textwidth]{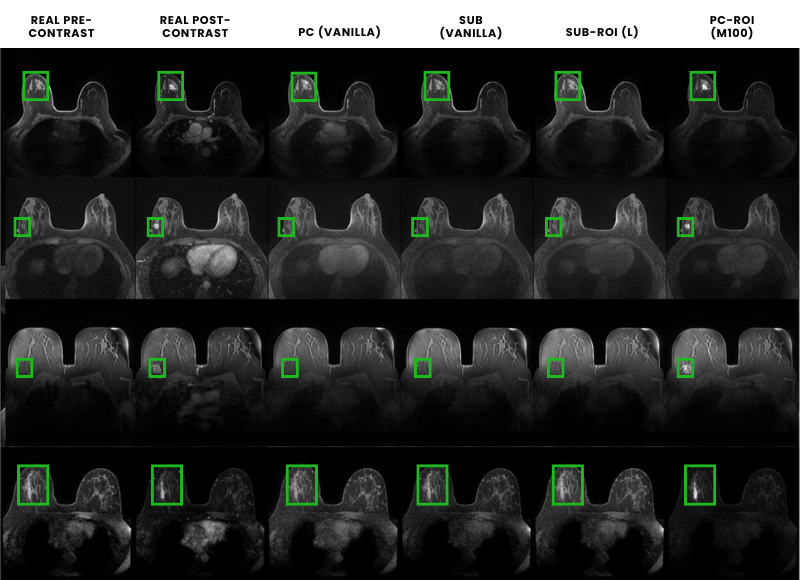}
    \caption{Qualitative synthetic test set samples comparing model variants with real pre- (col 1) and post-contrast (col 2) images. The variants include the vanilla post-contrast DDPM (col 3), the vanilla subtraction DDPM (col 4), ROI-aware loss subtraction DDPM (col 5), and ROI-mask conditioned post-contrast DDPM (col 6). The lesion area is highlighted using green bounding boxes.}
    \label{fig:4comparison}
\end{figure}

\begin{figure*}[ht]
\centering    
\includegraphics[width=0.9\textwidth]{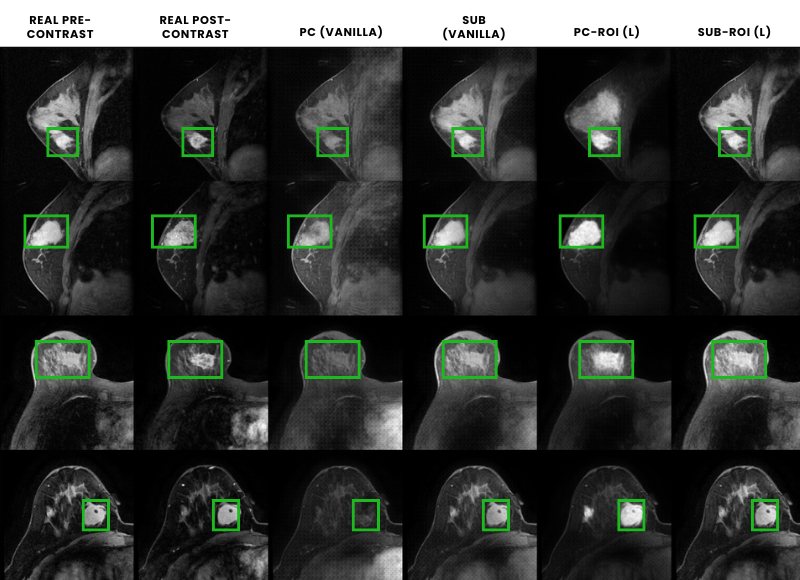}
\caption{Synthetic single-breast test set cases for visual comparison between model variants and real pre-contrast (col 1) and real post-contrast (col 2). The variants include the vanilla post-contrast (col 3) and subtraction DDPM model (col 4), ROI-aware loss post-contrast model (col 5), and ROI-aware loss subtraction model (col 6). The lesion area in each image is highlighted in a green bounding box.}
\label{fig:unifinalcomparison}
\end{figure*}

\section{Results}
\label{sec:results}

\subsection{Qualitative Evaluation and Comparison}

Figure~\ref{fig:4comparison} presents a visual comparison across model variants for four patient cases. While all models exhibit high-quality synthesis and realistic appearance, some limitations remain in accurately detecting lesions and rendering appropriate contrast. Compared to the vanilla post-contrast (PC) and subtraction (SUB) DDPMs, ROI-aware models, in particular, mask-conditioned models such as PC-ROI\textsubscript{(M100)} yield visibly improved enhancement within tumor regions with promising synthetic contrast patterns. Such mask-guided approaches are particularly relevant in applications with known approximate tumor locations, which include treatment response assessment, longitudinal surveillance or pre-treatment monitoring.
%
%
Training on full-breast images allows to learn patterns concerning anatomical symmetry relevant to tumor characterization. In contrast, single-breast models, while potentially benefiting from an enhanced focus on a single breast, on the other hand, can miss contralateral cues, which can be relevant in complex contrast enhancement 
and lesion detection cases. For qualitative single breast results, Fig. \ref{fig:unifinalcomparison} presents the same visual comparison as \ref{fig:4comparison} but in single-breast cases.
%
%
%
Notably, ROI-aware supervision commonly enhances boundary sharpness and better approximates the lesion intensity observed in real DCE-MRI scans. Overall, subtraction-based and ROI-guided variants exhibit stronger lesion reconstruction quality.

\begin{table*}[ht]
\centering
\caption{Quantitative results and ablations for \textit{full breast DCE-MRI synthesis} based on distribution-level metrics (FRD$\downarrow$, FID$\downarrow$) as well as paired image comparison metrics (MAE$\downarrow$, LPIPS$\downarrow$, SSIM$\uparrow$, PSNR$\uparrow$). Prefixes: PC=Post-Contrast, SUB=Subtraction. Suffixes: (100)=Model trained for 100 epochs rather than 50 epochs, (M)=Region-of-Interes(ROI)-masked, (L)=ROI-aware loss. PC models are compared to corresponding \textit{Real PC} and SUB models to corresponding \textit{Real SUB} images. Best values in \textbf{bold}.}
\label{tab:ablation_study_final}
\renewcommand{\arraystretch}{1.4}
\setlength{\tabcolsep}{5pt}
\resizebox{\textwidth}{!}{
\begin{tabular}{lcccccc}
\toprule
\textbf{Model Variant} & \textbf{FRD$\downarrow$} & \textbf{FID$\downarrow$} & \textbf{MAE$\downarrow$} & \textbf{LPIPS$\downarrow$} & \textbf{SSIM$\uparrow$} & \textbf{PSNR$\uparrow$} \\
\midrule
\multicolumn{7}{c}{\textbf{Full-Breast (metrics computed on full image)}} \\
\midrule
\textbf{Real Pre vs Real PC}     & \textbf{11.02} & \textbf{57.79}  & .062 ± .206 & \textbf{.193 ± .036} & .811 ± .021 & 23.09 ± 2.86 \\
PC\textsubscript{(Vanilla)}            & 59.10 & 74.49  & .072 ± .031 & .203 ± .033 & .815 ± .066 & 22.10 ± 3.51 \\
PC\textsubscript{(Vanilla100)}         & 58.85 & 72.56  & .071 ± .031 & .202 ± .033 & .816 ± .066 & 22.24 ± 3.55 \\
PC-ROI\textsubscript{(M)}          & 29.29 & 73.29  & .064 ± .026 & .203 ± .032 & .823 ± .063 & 23.30 ± 3.62 \\
PC-ROI\textsubscript{(M100)}       & 31.23 & 72.68  & .\textbf{060 ± .023} & .197 ± .032 & \textbf{.826 ± .063} & \textbf{23.94 ± 3.56} \\
PC-ROI\textsubscript{(L)}          & 23.45 & 75.13  & .065 ± .022 & .235 ± .043 & .819 ± .060 & 22.85 ± 3.01 \\
\hdashline
\textbf{Real Pre vs Real SUB}     & 100.46 & 165.91 & .862 ± .055 & .544 ± .034 & .028 ± .024 & 7.05 ± .45 \\
SUB\textsubscript{(Vanilla)}           & 28.99  & 69.75  & \textbf{.054 ± .021} & \textbf{.194 ± .037} & \textbf{.847 ± .061} & \textbf{25.63 ± 3.35} \\
SUB-ROI\textsubscript{(L)}         & \textbf{21.29}  & \textbf{67.07}  & \textbf{.054 ± .017} & .195 ± .035 & .826 ± .062 & 24.11 ± 2.75 \\
\midrule
\multicolumn{7}{c}{\textbf{Full-Breast (metrics computed on ROI)}} \\
\midrule
\textbf{Real Pre vs Real PC}       & 42.10 & 95.64  & .209 ± .054 & \textbf{.140 ± .061} & \textbf{.495 ± .191} & 16.94 ± 3.38 \\
PC\textsubscript{(Vanilla)}            & 59.10 & 129.10 & .334 ± .134 & .310 ± .098 & .303 ± .193 & 14.33 ± 3.49 \\
PC\textsubscript{(Vanilla100)}         & 68.85 & 140.69 & .389 ± .151 & .324 ± .097 & .283 ± .186 & 13.17 ± 3.41 \\
PC-ROI\textsubscript{(M)}          & 29.29 & 70.69  & .185 ± .066 & .221 ± .084 & .458 ± .182 & 19.01 ± 2.77 \\
PC-ROI\textsubscript{(M100)}       & 27.82 & \textbf{67.04}  & \textbf{.175 ± .057} & .215 ± .082 & .475 ± .171 & \textbf{19.28 ± 2.38} \\
PC-ROI\textsubscript{(L)}          & \textbf{21.29} & 79.06  & .265 ± .104 & .258 ± .090 & .302 ± .182 & 16.34 ± 3.02 \\
\hdashline
\textbf{Real Pre vs Real SUB}      & 211.29 & 210.50 & .391 ± .052 & .388 ± .063 & .326 ± .207 & 8.66 ± 3.24 \\
SUB\textsubscript{(Vanilla)}           & 29.44 & 98.37  & .301 ± .133 & .281 ± .095 & .342 ± .207 & 15.32 ± 3.67 \\
SUB-ROI\textsubscript{(L)}         & \textbf{23.45} & \textbf{89.48}  & \textbf{.247 ± .107} & \textbf{.265 ± .094} & \textbf{.374 ± .210} & \textbf{16.84 ± 3.42} \\

\bottomrule
\end{tabular}
}
\end{table*}

\subsection{Quantitative Assessment and Benchmarking}

We quantitatively evaluate our model variants calculating image synthesis evaluation metrics for both full image and for the tumor-containing region-of-interest (ROI) of each image. Apart from benchmarking PC and SUB image synthesis, we also comparatively assess full breast MRI and single-breast image synthesis depicted in Table~\ref{tab:ablation_study_final} and Table~\ref{tab:ablation_study_final_single_breast}, respectively. To this end, we apply mutually-complementary \cite{osuala2025simulating} image quality metrics, namely, Mean Absolute Error (MAE), Structural Similarity Index (SSIM)~\cite{wang2004image}, Peak Signal-to-Noise Ratio (PSNR), Learned Perceptual Image Patch Similarity (LPIPS)~\cite{zhang2018unreasonable}, Fréchet Inception Distance (FID)~\cite{heusel2017gans} and Fréchet Radiomics Distance (FRD) ~\cite{osuala2024towards,konz2025frechetradiomicdistancefrd}.
\begin{table*}[ht]
\centering
\caption{Quantitative results and ablations for single-breast image synthesis based on the same metrics and prefixes as in Table \ref{tab:ablation_study_final}. 
}
\label{tab:ablation_study_final_single_breast}
\renewcommand{\arraystretch}{1.4}
\setlength{\tabcolsep}{5pt}
\resizebox{\textwidth}{!}{
\begin{tabular}{lcccccc}
\toprule
\textbf{Model Variant} & \textbf{FRD$\downarrow$} & \textbf{FID$\downarrow$} & \textbf{MAE$\downarrow$} & \textbf{LPIPS$\downarrow$} & \textbf{SSIM$\uparrow$} & \textbf{PSNR$\uparrow$} \\
\midrule
\multicolumn{7}{c}{\textbf{Single-Breast (metrics computed on full image)}} \\
\midrule
\textbf{Real Pre vs Real PC}       & \textbf{11.02} & \textbf{79.89}  & .120 ± .053 & \textbf{.257 ± .057} & .634 ± .126 & \textbf{20.82 ± 3.09} \\
PC\textsubscript{(Vanilla)}            & 177.10 & 82.29  & .128 ± .047 & .294 ± .060 & .700 ± .097 & 18.87 ± 2.87 \\
PC-ROI\textsubscript{(L)}          & 60.74  & 104.03 & \textbf{.111 ± .041} & .304 ± .068 & \textbf{.702 ± .096} & 20.34 ± 3.02 \\
\hdashline
\textbf{Real Pre vs Real SUB}      & 107.62 & 131.16 & .736 ± .101 & .514 ± .030 & .047 ± .054 & 8.12 ± 1.01 \\
SUB\textsubscript{(Vanilla)}           & 59.82  & 88.10  & \textbf{.092 ± .036} & .266 ± .061 & \textbf{.715 ± .095} & \textbf{22.52 ± 2.80} \\
SUB-ROI\textsubscript{(L)}         & \textbf{43.55}  & \textbf{80.47}  & .105 ± .045 & \textbf{.249 ± .052} & .695 ± .099 & 21.89 ± 2.97 \\
\midrule
\multicolumn{7}{c}{\textbf{Single-Breast (metrics computed on ROI)}} \\
\midrule
\textbf{Real Pre vs Real PC}       & 71.02 & \textbf{79.89}  & \textbf{.126 ± .154} & \textbf{.257 ± .057} & \textbf{.634 ± .126} & \textbf{20.82 ± 3.09} \\
PC\textsubscript{(Vanilla)}            & 177.10 & 210.50 & .575 ± .227 & .499 ± .092 & .172 ± .167 & 10.17 ± 3.39 \\
PC-ROI\textsubscript{(L)}          & \textbf{60.74}  & 151.72 & .340 ± .138 & .438 ± .104 & .187 ± .181 & 14.31 ± 3.13 \\
\hdashline
\textbf{Real Pre vs Real SUB}      & 85.35 & \textbf{101.60} & .321 ± .107 & .368 ± .081 & \textbf{.288 ± .235} & 13.68 ± 3.44 \\
SUB\textsubscript{(Vanilla)}           & 59.82  & 136.59 & .307 ± .161 & .402 ± .109 & .256 ± .207 & 15.44 ± 3.95 \\
SUB-ROI\textsubscript{(L)}         & \textbf{43.55}  & 104.60 & \textbf{.266 ± .125} & \textbf{.366 ± .101} & .278 ± .205 & \textbf{16.53 ± 3.44} \\
\bottomrule
\end{tabular}
}
\end{table*}
The results demonstrate the ability of diffusion models to successfully synthesize both PC and SUB images from pre-contrast inputs in both full-breast and single-breast settings. 
When comparing vanilla models for PC and SUB image synthesis while considering their respective \textit{real pre vs real target} baselines, particularly the SUB\textsubscript{(Vanilla)} model improves considerably over its baseline to this end outperforming its PC counterpart. This observation holds true for both full-image as well as ROI-based metric computation across distribution-based and image-level metrics. This trend of SUB outperforming PC synthesis is similarly observable for both full-breast MRI slice models and cropped single-breast image synthesis models. 
Comparing such subtraction based single and full-breast models, the full-breast models more consistently outperformed the respective \textit{real pre vs real SUB} baseline across metrics than their single-breast counterparts for across both synthetic ROI and full synthetic image evaluation metrics. For full breast MRI slice generation, the inclusion of a tumor-aware loss (e.g. in SUB-ROI\textsubscript{(L)}) generally further improved the synthesis performance for ROI-based metrics, while only resulting in improvements on distribution leverl (FRD, FID) for full image metrics (compared to SUB\textsubscript{(Vanilla)}). In single breast generation models, the same trend is observable for both ROI-based metrics and full-image metrics, with the exception of LPIPS in the latter, where single-breast SUB-ROI\textsubscript{(L)} (.249 ± .052) outperforms SUB\textsubscript{(Vanilla)} (.266 ± .061).


\subsection{Reader Study: Results and Evaluation}

\begin{figure}[ht]
    \centering
    \includegraphics[width=0.95\textwidth]{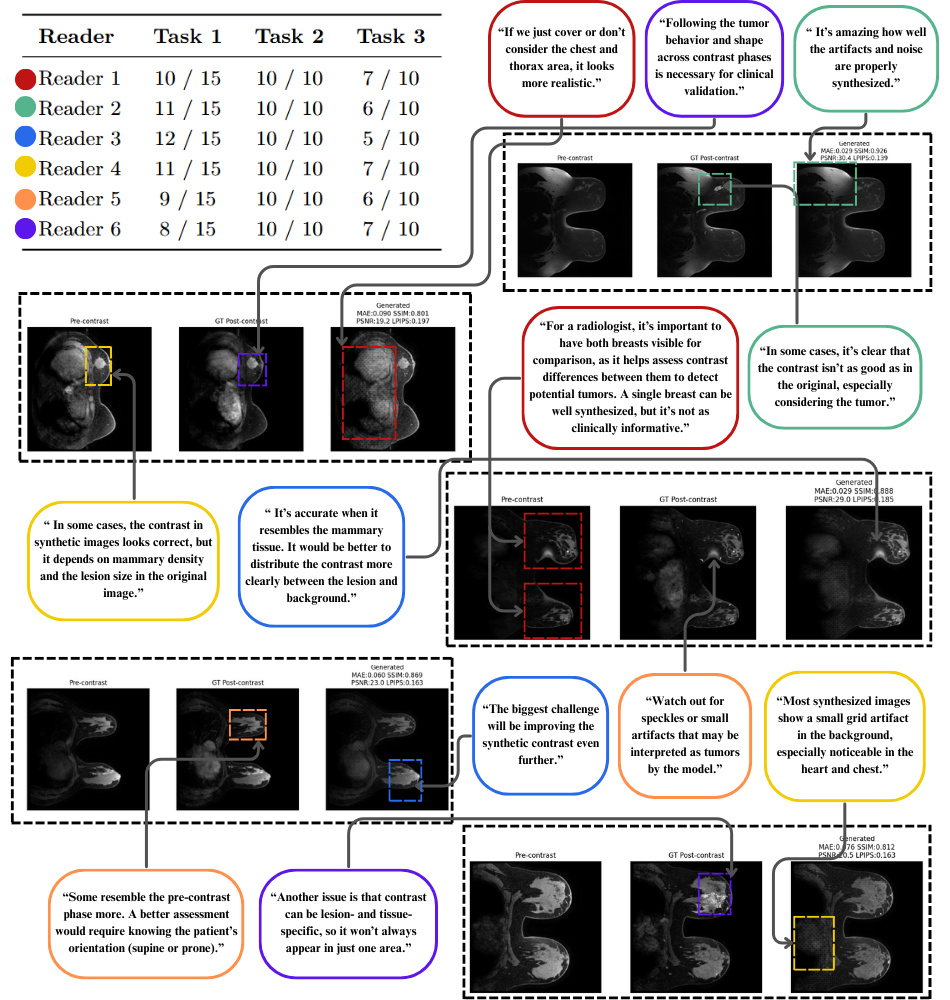}
    \caption{Selection of expert reader annotations and respective remarks acquired during the assessment and discussion of synthetic images in Tasks 1, 2 and 3.}
    \label{fig:reader-study}
\end{figure}

For reader assessment, all three tasks were based on outputs of (SUB\textsubscript{(Vanilla)}) due to its promising performance compared to (PC\textsubscript{(Vanilla)}), with additional comparison of ROI-aware and ROI-masked model variants.
Task 1 showed moderate difficulty, with readers achieving between 60\% and 80\% accuracy. In Task 2, the readers were able to consistently select the real post-contrast image 
confirming consistent perceptual differences, but also noting having learned from feedback of their per-image performance in Task 1. Subjective realism scores in Task 3 ranged from 5 to 7 out of 10.
The readers' annotations highlighted several aspects: Readers acclaimed the anatomical fidelity of the synthetic images, especially in breast parenchyma, skin boundaries, and overall mammary tissue structure. Some experts noted the effectiveness of artifact simulation and the preservation of coarse contrast patterns in ROI-guided variants.
However, limitations were also noted such as insufficient tumor enhancement, particularly during early contrast phases, flat background appearance, grid-like artifacts, and minor asymmetries in the heart and chest regions. These elements were frequently used by experts to distinguish synthetic from real images. The readers emphasized the clinical potential of the generative modeling approach and viewed the ROI-aware and ROI-masked variants as promising for specific tasks such as treatment monitoring, response assessment, or as adjunct tools in abbreviated MRI protocols, where full contrast dynamics can be unavailable or undesired. Selected expert annotations are visualized in Fig. \ref{fig:reader-study}.

\section{Discussion and Conclusions}
\label{sec:discussion}

We proposed and tested pre-contrast-conditioned DDPM frameworks where 
our subtraction-based DDPMs (SUB\textsubscript{(Vanilla)}) quantitatively outperformed their post-contrast counterparts (PC\textsubscript{(Vanilla)}) based on respective baseline comparisons. In subtraction variants the synthesis task is likely simplified by allowing the model to focus exclusively on the residual of the contrast signal, 
while most of the anatomy remains static. ROI-aware losses improved quantitative results while not requiring segmentation masks at inference, thus enabling lesion detection applications. For scenarios where lesion localization is known a priori, e.g. in cancer treatment, segmentation mask conditioned models 
notably improve qualitative ROI results. 
%
%
Expert reader evaluation confirmed 
anatomical realism, especially for subtraction-based and ROI-conditioned models. While some cases showed limited synthetic contrast or synthetic artifacts (e.g., grid textures), ROI-guided variants were perceived as more clinically promising, particularly for 
monitoring or abbreviated protocols.
%
Future work will extend the scope to include ablation of individual loss components, the application of losses directly on subtraction images (in addition to PC reconstructions), and an analysis of optimal loss weighting strategies within the total loss formulation.
Further promising avenues include 3D synthesis, modeling of DCE-MRI temporal dynamics \cite{osuala2024towards,schreiter2024virtual,lang2025temporalneuralcellularautomata}, and comparison of image-level DDPMs to their latent counterparts \cite{rombach2022high} including exploration of conditioning and fine-tuning strategies. 
Overall, our promising DCE-MRI synthesis results motivate further investigation into their clinical applicability for safe, rapid, and scalable imaging with reduced reliance on contrast agents, especially in patients where their use is contraindicated.



\section*{Disclosure of Interests}

The authors declare no conflict of interest.
Some authors are affiliated with Siemens Healthineers. This work reflects their personal scientific views and does not represent the company’s official position or endorsement.

%
%
%
\bibliographystyle{splncs04}
\bibliography{mybibliography}
\end{document}